\documentclass[10pt]{article}
\usepackage{epsf}

\newlength{\dinwidth}
\newlength{\dinmargin}
\def\lapproxeq{\lower .7ex\hbox{$\;\stackrel{\textstyle <}{\sim}\;$}}
\def\gapproxeq{\lower .7ex\hbox{$\;\stackrel{\textstyle >}{\sim}\;$}}

\def\be{\begin{equation}}
\def\ee{\end{equation}}
\def\bea{\begin{eqnarray}}
\def\eea{\end{eqnarray}}

\def\gtrsim{{ \;\raisebox{-.7ex}{$\stackrel{\textstyle >}{\sim}$}\; }}
\def\lesim{{ \;\raisebox{-.7ex}{$\stackrel{\textstyle <}{\sim}$}\; }}

\def\GeV{{\rm GeV}}

\def\bb{{b\bar{b}}}
\newcommand{\eq}[1]{(\ref{eq:#1})}

\begin{document}
\titlepage

\begin{flushright}
IPPP/03/51 \\
DCPT/03/102 \\
4 December 2003 \\
\end{flushright}

\vspace*{4cm}

\begin{center}
{\Large \bf Extending the study of the Higgs sector at the LHC by proton tagging}

\vspace*{1cm} \textsc{A.B.~Kaidalov$^{a,b}$, V.A.~Khoze$^{a,c}$, A.D. Martin$^{a,d}$ and M.G. Ryskin$^{a,c}$} \\

\vspace*{0.5cm} $^a$ Department of Physics and Institute for
Particle Physics Phenomenology, \\
University of Durham, DH1 3LE, UK \\
$^b$ Institute of Theoretical and Experimental Physics, Moscow, 117259, Russia\\
$^c$ Petersburg Nuclear Physics Institute, Gatchina,
St.~Petersburg, 188300, Russia \\
$^d$ Department of Physics, University of Canterbury, Christchurch, New Zealand \\
\end{center}

\vspace*{1cm}

\begin{abstract}
We show that forward proton tagging may significantly enlarge the potential of  studying the Higgs sector at the
LHC. We concentrate on Higgs production via central exclusive diffractive processes (CEDP). Particular attention
is paid to regions in the MSSM parameter space where the partial width of the Higgs boson decay into two gluons
much exceeds the SM case; here the CEDP are found to have special advantages.
\end{abstract}

\section{Introduction}

The Higgs sector is (so far) an elusive basic ingredient of the fundamental theory of particle interactions.
Searches  for Higgs bosons, and the study of their properties, are one of the primary goals of the Large Hadron
Collider (LHC) at CERN, which is scheduled to commence taking data in the year 2007. The conventional folklore is
that (under reasonable model assumptions) at least one Higgs boson should be discovered at the LHC. In particular,
if the light Higgs predicted by the Standard Model (SM) exists\footnote{The
theoretical progress in the calculations of the signal and background processes has been enormous in
 recent years, reaching an accuracy of the next-to-next-to-leading QCD corrections for some relevant cross
sections, see for example \cite{AD} and references therein. At the same time, experimentally related issues of the
Higgs search physics have been thoroughly investigated, see for example \cite{CMS}.},
it will almost certainly be found at the LHC in the first years of running or even maybe before, at the Tevatron.
Moreover the LHC should provide a complete coverage of the SM Higgs mass range.

However there is a strong belief that the Standard Model, in its minimal form with a single Higgs,  cannot be the
fundamental theory of particle interactions. Various extended models predict a large diversity of Higgs-like
bosons with different masses, couplings and even CP-parities. The most elaborated extension of the Standard Model
is, currently, the Minimal Supersymmetric Standard Model (MSSM) in which there are three neutral ($h$, $H$ and
$A$) and two charged ($H^+,H^-$) Higgs bosons, where $h$ and $H$ are CP-even ($m_h< m_H$) and $A$ is CP-odd. Just
as for the Standard Model, this benchmark SUSY model has been studied in great detail; for a recent review
see~\cite{CH}. In the MSSM, the properties of the Higgs sector are characterized by the values of two independent
input parameters, typically chosen to be the pseudoscalar Higgs boson mass, $m_A$, and the ratio, $\tan\beta$, of
the vacuum-expectation-values of the two Higgs doublet fields. At tree level, the pseudoscalar $A$ does not couple
to the gauge bosons and its couplings to down- (up-) type fermions are (inversely) proportional to $\tan\beta$.
Within the MSSM, the mass of the $h$-boson is bounded to $m_h \lesim 135$~GeV (see, for example, \cite{hhw} and
references therein), while the experimental 95\% CL lower limit for a neutral scalar Higgs is $m_h \simeq m_A  \sim
92$~GeV~\cite{LEP}.

Beyond the Standard Model the properties of the neutral Higgs bosons can differ drastically from SM
expectations. In some extended models, the limit for a neutral Higgs can go down to below 60~GeV. This occurs, for
instance in the MSSM with explicit CP-violation, where the mass eigenstates of the neutral Higgs bosons do not
match the CP eigenstates $h,H,A$, see, for example,~\cite{CPX}. Further examples are the models with extra
dimensions where a Higgs can mix with the graviscalar of the Randall--Sundrum scenario~\cite{RS}, which is called
the radion (see for example \cite{DGGT}). In the latter case,  due to the trace anomaly, the gluon--gluon coupling
is strongly enhanced, which makes this graviscalar especially attractive for searches in gluon-mediated processes,
assuming that the background issues can be overcome.  These extended scenarios would complicate the study of the
neutral Higgs sector using the conventional (semi)inclusive strategies.

After the discovery of a Higgs candidate the immediate task will be to establish its quantum numbers, to verify
the Higgs interpretation of the signal, and to make precision measurements of its properties. The separation and
identification of different Higgs-like states will be especially challenging. It will be an even more delicate
goal to establish the nature of a newly-discovered heavy resonance state. For example, how can one discriminate
between the Higgs of the extended SUSY sector from the graviscalar of the Randall--Sundrum scenario (or, even
worse, from a mixture of the Higgs and the radion). As was shown in \cite{KKMRCentr}, the central exclusive
diffractive processes (CEDP) at the LHC can play a crucial role in solving these problems, which are so vital for
the Higgs physics.  These processes are of the form
\begin{equation}
pp\to p + \phi + p, \label{eq:cedp}
\end{equation}
where the $+$ signs denote the rapidity gaps on either side of the Higgs-like state $\phi$.  They have unique
advantages as compared to the traditional non-diffractive approaches~\cite{KMRProsp,DKMOR}. These processes allow
the properties of $\phi$ to be studied in an environment in which there are no secondaries from the underlying
events. In particular, if the forward protons are tagged, then the mass of the produced central system $\phi$ can
be measured to high accuracy by the missing mass method.  Indeed, by observing the forward protons, as well as the
$\phi\to\bb$ pairs in the central detector, one can match two simultaneous measurements of the $\phi$ mass:
$m_\phi = \Delta m_{\rm missing}$ and $m_\phi = m_\bb$. Moreover, proton taggers allow the damaging effects of
multiple interactions per bunch crossing (pile-up) to be suppressed and hence offer the possibility of studying
CEDP processes at higher luminosities~\cite{DKMOR}.  Thus the prospects of the precise mass determination of the
Higgs-like states, and even of the direct measurements of their widths and $\phi\to\bb$ couplings, looks feasible
using these processes. A promising option is to use the forward proton taggers as a spin-parity
analyser~\cite{KKMRCentr}. This may provide valuable additional (and in some cases unique) information in the
unambiguous identification of a newly discovered state.

In Section~\ref{sec:disoverypotential} we illustrate the advantages of CEDP,~(\ref{eq:cedp}), in exploring the
Higgs sector in specific parameter ranges of the MSSM.  First in Section~\ref{sec:icr} we consider the so-called
``intense-coupling'' regime~\cite{BDMV,BDN} where the $\gamma\gamma$, $WW^\star$ and $ZZ^\star$ Higgs decay modes
are suppressed but where, on the other hand, the CEDP cross section is enhanced, in comparison with the SM.  Next,
in Section~2.2, we discuss the decoupling limit ($m_A>2M_Z$, $\tan\beta>5$) where the light scalar, $h$, looks
very similar to the SM Higgs.  In this case the discovery of the heavier scalar $H$ is crucial to establish the
underlying structure of the electroweak-symmetry-breaking dynamics. If $m_H<250$~GeV, then the $H$ boson may be
observed and studied in CEDP.  In Section~2.3 we show that CEDP may cover the ``windows'' where, once the $h$
boson is discovered, it is not possible to disentangle the $H$ or $A$ bosons by traditional means.
In Section~3 we explain how studies of CEDP may help to determine the properties of the Higgs-like states, once
they have been discovered.

The plan of the paper is first to present the predictions and then, in Section~4, to explain how the cross
sections for the CEDP of the $\phi$ CP-even and CP-odd Higgs states are calculated. In this Section we also
enumerate the uncertainties in the predictions.  Simple formulae to approximate the CEDP cross sections for $pp\to
p + \phi + p$ with $\phi\to\bb$, and the corresponding QCD $\bb$ background, are given in Section~5.  These allow
quick, reliable estimates to be made of the signal-to-background ratio for a Higgs of any mass, and both CP
parities.

In Section~6 we focus attention on ways to identify the pseudoscalar boson $A$.   It is unlikely
that the pure exclusive diffractive production process will have sufficient events, and so possible semi-inclusive
signals are proposed and studied.

\section{Discovery potential of diffractive processes}  \label{sec:disoverypotential}

The main aim of this paper is to show how the Higgs discovery potential at the LHC may be enlarged by using the
unique advantages of forward proton tagging. Special attention is paid to the cases where the partial widths for
the decay of Higgs bosons into two gluons greatly exceed the SM values. As mentioned above, we shall illustrate
the advantages of using forward proton tagging to study the Higgs sector by giving predictions for a few examples
within the MSSM model.

\subsection{The intense coupling regime}  \label{sec:icr}

The  ``intense-coupling regime''  \cite{BDMV,BDN} is where the masses of all three neutral Higgs bosons are close
to each other and the value of $\tan\beta$ is large. Then the $\gamma\gamma,WW^\star,ZZ^\star$ decay modes (which
are among the main detection modes for the SM Higgs)  are strongly suppressed. This is the regime where the variations
of all MSSM Higgs masses and couplings are very rapid. This region is considered as one of the most challenging
for the (conventional) Higgs searches at the LHC, see \cite{BDN} and references therein.  On the other hand, here
the CEDP cross sections are enhanced by more than an order of magnitude (due to the large $gg\to \phi$ couplings).
Therefore the expected significance of the CEDP signal becomes quite large.  Indeed, this is evident from Fig.~1,
which shows the cross sections for the CEDP production of $h,H,A$ bosons as functions of their mass for
$\tan\beta=30$ and 50.
\begin{figure}[htb]
\begin{center}
\centerline{\epsfxsize=10cm\epsfbox{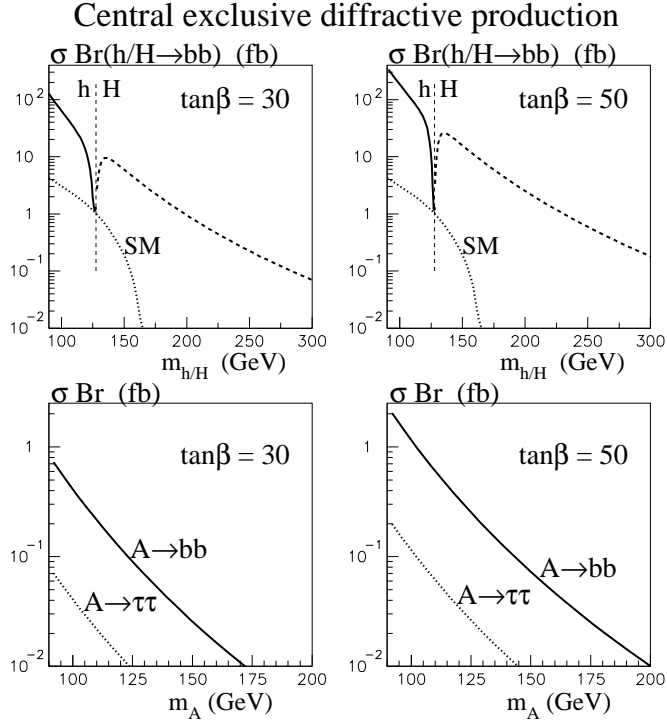}} \caption{The cross sections, times the appropriate $\bb$ and
$\tau^+\tau^-$ branching fractions, predicted for central exclusive diffractive production of $h(0^+)$, $H(0^+)$
and $A(0^-)$ MSSM Higgs bosons (for $\tan\beta=30$ and 50) at the LHC, obtained using the MRST99~\cite{MRST99}
gluon distribution. The dotted curve in the upper plots shows the cross section for the production of a SM Higgs
boson. The vertical line separates the mass regime of light $h(0^+)$ and heavy $H(0^+)$ bosons.
The curves were computed using version 3.0 of the HDECAY code~\cite{HDEC}, with all other
parameters taken from Table 2 of~\cite{HDEC}; the radiative corrections were included
according to Ref.~\cite{HHW}. \label{fig:a}}
\end{center}
\end{figure}
The cross sections are computed as described in Section~5, with the widths and properties of the Higgs scalar
($h,H$) and pseudoscalar ($A$) bosons obtained from the HDECAY code, version 3.0~\cite{HDEC}, with all other
parameters taken from Table~2 of \cite{HDEC}; also we take ${\rm IMODEL} = 4$, which means the radiative
corrections are included according to Ref.~\cite{HHW}.

Let us focus on the main $\phi\to \bb$ decay mode\footnote{The exclusive diffractive studies in Ref.~\cite{DKMOR}
addressed mainly the $\bb$ decay mode.  The use of the $\tau\tau$ decay mode requires an evaluation of the $pp\to
p + \tau\tau + p$ background, especially of the possibility of misidentifying gluon jets as $\tau$'s in the CEDP
environment. Currently, for inclusive processes, the probability, $P(g/\tau)$, to misidentify a gluon as a $\tau$
is estimated to be about 0.002~\cite{CKN}.}.
This mode is well suited for CEDP studies, since a $P$-even, $J_z=0$ selection rule~\cite{Liverpool,KMRItal,KMRmm}
suppresses the QCD $\bb$ background at LO. Indeed, at the LHC, the signal-to-background ratio is $S/B\sim3$ for a
Standard Model Higgs boson of mass $m=120$~GeV, if the experimental cuts and efficiencies quoted in \cite{DKMOR}
are used. This favourable ratio was obtained since it was argued that proton taggers can achieve a missing mass
resolution of $\Delta m_{\rm missing}\simeq 1$~GeV.  Due to the enhancement of the production of MSSM scalars, for
large $\tan\beta$, the QCD background becomes practically negligible.
Note that, normally, to estimate  the statistical significance
of the measurement of the signal cross section, we use the conservative
formula $\sqrt{S+B}$ for the statistical error. However
when we are looking for a new particle, which gives some peak
on the top of the background, the statistical error may be much lower.
It is given just by the fluctuation of the background --
$\sqrt{B}$. On the other hand for a low number of events, it is appropriate
to use a Poisson, rather than a Gaussian, distribution.
Thus the statistical significance
to discover a new particle has to be evaluated more precisely.
For instance, with the cuts and efficiencies quoted in [11],
we estimate that  the QCD background is $B$ =  40(4) events
for an integrated luminosity of ${\cal L} = 300~{\rm fb}^{-1}$
(30~fb$^{-1}$), if we take $M = 120$ GeV and $\Delta M = 1$ GeV.
Thus, to obtain a confidence coefficient of $5.7\times 10^{-7}$ (equivalent to a
statistical significance  of $5\sigma$ for a Gaussian
distribution\cite{RPP}),  it is sufficient for the cross section
of a Higgs signal, times the $\bb$ branching fraction, to satisfy

\begin{equation}
{\rm Br}(\bb)\cdot\sigma > 0.7~{\rm fb}~(2.7~{\rm fb}) \label{eq:cc1}
\end{equation}
if we take the integrated luminosity to be ${\cal L} = 300~{\rm fb}^{-1}$
(30~fb$^{-1}$).

As can be seen from Fig.~1, for MSSM with $\tan\beta=50$  the expected CEDP
cross section times branching fraction, ${\rm Br}(H\to\bb)\sigma_H$, is greater than 0.7fb for
masses up to $m_H\sim 250$ GeV.
The situation is worse for pseudoscalar, $A$, production, since the cross section is suppressed by the
$P$-even selection rule.  Thus the CEDP filters out pseudoscalar production, which allows the possibility to study
pure $H$ production, see Fig.~1.  This may be also useful in the decoupling limit, to which we now turn.

\subsection{The decoupling limit} \label{sec:decouplinglimit}

For $m_A > 2m_Z$ and $\tan\beta > 5$, there is a wide domain in MSSM parameter space where the light scalar $h$
becomes indistinguishable from the SM Higgs, and the other two neutral Higgs states are approximately degenerate
in mass. This is the so-called decoupling limit~\cite{HEH}.  Moreover in many theories with a non-minimal Higgs
sector, there are significant regions of parameter space that approximate the decoupling limit; see, for instance,
\cite{CH}. In this regime the discovery of the heavier non-minimal Higgs scalar $H$ is crucial for establishing
the underlying structure of the electroweak-symmetry-breaking dynamics. Here, forward proton tagging can play an
important role in searching for (at least) the $H$-boson, if it is not too heavy ($m_H \lesim 250$~GeV). For large
values of $\tan\beta$ the decoupling regime essentially starts at $m_A\simeq 170$~GeV. As can be seen in Fig.~1,
the cross section is still sufficiently large to ensure the observation of the $H$ boson up to $m_A\simeq250$~GeV.
To illustrate the moderate $\tan\beta$ region we consider the case $m_A=185$~GeV and $\tan\beta = 7$.  Then the
CEDP cross section is
\begin{equation}
{\rm Br}(H\to\bb)\sigma_H = 0.17~{\rm fb}.\label{eq:new1}
\end{equation}
Thus for an integrated luminosity of ${\cal L} = 300~{\rm fb}^{-1}$ we have about 8.5 events, after applying the
experimental cuts and efficiencies discussed in Ref.~\cite{DKMOR}. The corresponding background is about
2.6~events. Using again a Poisson distribution, as we did in the case of (\ref{eq:cc1}), then the expected
statistical significance to observe such an $H$ boson is about $4\sigma$.  So detection may be feasible.

The possibility to use exclusive diffractive processes to explore larger masses will depend on various
experiment-related factors. In particular, on the prospects to achieve better mass resolution, $\Delta m$, at
higher mass, $m$.

\subsection{The `window' or `hole' regions} \label{sec:window}

There are regions of MSSM parameter space where, after the $h$ boson is discovered, it will be difficult to
identify the heavy $H$ or $A$ bosons by conventional non-diffractive processes.  Can the CEDP processes help here?
First consider the `notorious hole' in parameter space (sometimes called the ``LHC wedge''), defined by $m_A
\gtrsim 200$~GeV and $\tan\beta \sim\:$4--7~\cite{CH}. In this region, only the lightest Higgs $h$ can be
discovered even with integrated luminosity of the order of 300~fb$^{-1}$. Its properties are nearly
indistinguishable from those of a SM Higgs. Unfortunately here, even for $H$, the predicted CEDP cross section is
too low, assuming the efficiencies and cuts of Ref.~\cite{DKMOR}; see the heavy continuous curve in Fig.~2.
\begin{figure}[htb]
\begin{center}
\centerline{\epsfxsize=10cm\epsfbox{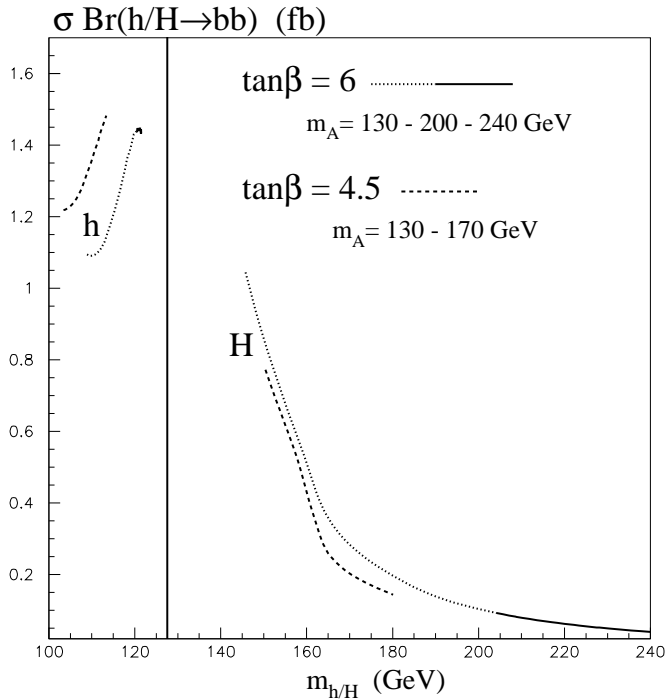}} \caption{The cross sections for the central exclusive diffractive
production of $h$ and $H$ bosons, $pp\to p + (h,H) + p$, times the appropriate $\bb$ branching fraction, as a
function of their mass, in two regions of MSSM parameter space where observation
of the $H$ using non-diffractive processes
is difficult.  Note that in the first region the curve changes from dotted to continuous when the mass of the
$A(0^-)$ bosons reaches 200~GeV; and that the value of $m_h$ is practically unchanged for values of $m_A$ in the
region $m_A>200$~GeV. \label{fig:b}}
\end{center}
\end{figure}

However, with the present understanding of the capabilities of the detectors it is plausible that similar windows
will occur in the interval $130 \lesim m_A\lesim 170$~GeV for $\tan\beta \simeq\:$4--6.\footnote{We are  grateful
to Sasha Nikitenko for pointing out the importance of the additional (nonconventional) studies in this region.}
That is, it may be impossible to identify the $H$ scalar by conventional processes at the $5\sigma$ confidence
level with 300~fb$^{-1}$ of combined ATLAS+CMS luminosity. From the predictions shown by the dashed line in
Fig.~2, we see that the part of this `hole', up to $m_H\sim 160$~GeV, may be covered by CEDP.  For illustration,
consider the case $m_H = 150$~GeV and $\tan\beta=4.5$.  With an integrated luminosity of ${\cal L} = 300~{\rm
fb}^{-1}$, the CEDP cross section of
\be {\rm Br}(H\to\bb)\sigma_H = 0.8~{\rm fb} \ee
gives about 40 events, after applying the experimental cuts and efficiencies discussed in Ref.~\cite{DKMOR}. The
corresponding background is about 10~events.  Thus, if we use a Poisson distribution, as we did in the case of
(\ref{eq:cc1}), then the expected statistical significance to observe such an $H$ boson will exceed $10\sigma$.

\subsection{Exotic scenarios}

Recently some interesting applications of the CEDP to studies of the more exotic Higgs scenarios, such as models
with CP violation and searches for the radions, have been discussed~\cite{CFLP}. However the feasibility of
applying these studies will require further critical discussion of the background issues.

\section{Potential of diffractive processes after a Higgs discovery} \label{sec:adp}

As was discussed in Refs.~\cite{DKMOR,KKMRCentr}, the use of CEDP, with forward proton tagging, can provide a
competitive way  to detect a Higgs boson. This approach appears to have special advantages for searches of the
CP-even Higgses of MSSM in the intense-coupling regime. Indeed, the CEDP cross section can exceed the SM result by
more than an order of magnitude, see Fig.~1. Once a Higgs-like signal is observed, it will be extremely important
to verify that it is indeed a Higgs boson. Here we describe where forward proton tagging can provide valuable help
to establish the nature of the newly discovered state and to verify whether it can be interpreted as a Higgs
boson. We also emphasize that  CEDP are well suited to study specific cases which pose difficulties for probing
the Higgs sector at the LHC.

It is generally believed that a detailed coverage of the Higgs sector will require complementary research
programmes at the LHC and a Linear $e^+e^-$ collider. Interestingly, in some sense, the implementation of forward
proton taggers allows some studies which otherwise would have to await the construction of a Linear $e^+ e^-$
collider. Moreover, in addition to the other advantages, forward proton tagging offers the possibility to study
New Physics in the photon--photon collisions, see for example~\cite{OWZ,Piotrzk,KMRPhot,KMRProsp}.

As mentioned above, if a candidate Higgs signal is detected it will be a challenging task to prove its Higgs
identity. Unlike the conventional inclusive approaches, the very fact of seeing the resonance state in CEDP
automatically implies that the newly discovered state  has the following fundamental properties. It must have zero
electric charge and be a colour singlet. Furthermore, assuming $P$ and $C$ conservation, the dominantly produced
state has a positive natural parity, $P=(-1)^J$ and even CP. Recall that the installation of forward proton
taggers also allows the attractive possibility of a spin-parity analysis. This may provide valuable additional
(and in some cases unique) leverage in establishing the origin of the discovered candidate state. In particular,
assuming CP conservation, the CEDP allow the $0^-,1^-,1^+$ states to be filtered out, leaving only an ambiguity
between the $0^{++}$ and $2^{++}$ states. Though without further efforts the $2^{++}$ state  cannot  be ruled out,
this would not be the most likely choice.

As discussed in \cite{KKMRCentr}, studying of the azimuthal correlations of the outgoing protons can allow further
spin-parity analysis. In particular, it may be possible to isolate the $0^-$ state. The azimuthal distribution
distinguishes between the production of scalar and pseudoscalar particles~\cite{KKMRCentr}. Note that with the
forward protons we can determine the CP-properties of the Higgs boson irrespective of the decay mode. Moreover,
CEDP allow the observation of the interference effects between the CP-even and CP-odd $gg\to\phi$ transitions (see
\cite{KMRInt..}).

Following the discovery, and the determination of the basic quantum numbers, of the new state, the next step is to
check whether the coupling properties match those expected for the Higgs boson. For this, we need to detect the
signal in at least two different channels. The CEDP mechanism provides an excellent opportunity
\footnote{Recall that for more inclusive processes the signal is overwhelmed by the QCD $\bb$ background.}
to detect the $\bb$ mode. The other decay channel which is appropriate for CEDP studies is the $\tau \tau$ decay
mode (or $WW^\star$, or $\mu \mu$, if the event rate is sufficient).  These measurements allow a non-trivial check
of the theoretical models.  For example, the precise measurement of the partial decay widths into  tau and bottom
quark pairs can provide a way to discriminate between supersymmetric and non-supersymmetric
Higgs sectors, and to probe CP-violating effects (for recent papers see, for
instance,~\cite{IN} and references therein).

For illustration, consider the following topical example, namely the intense coupling limit of MSSM. As mentioned
before, this regime is especially difficult for conventional inclusive studies of the MSSM Higgs~\cite{BDMV,BDN}.
On the other hand, with forward proton taggers, CEDP offer an ideal probe.

For numerical purposes, we choose the same parameters as in \cite{BDN}, in particular, $\tan\beta=30$, together
with results for $\tan\beta=50$. The salient features of this regime are that there is almost a mass degeneracy of
the neutral Higgs states, and their total widths can be quite large (due to the $\tan^2\beta$ increase) and  reach
up to 1--2~GeV. This can be seen from Fig.~3, where we plot the mass differences $\Delta M = m_A - m_h$ and
$\Delta M = m_H - m_A$, and the widths of the bosons, as a function of the pseudoscalar mass $m_A$.
\begin{figure}[htb]
\begin{center}
\centerline{\epsfxsize=10cm\epsfbox{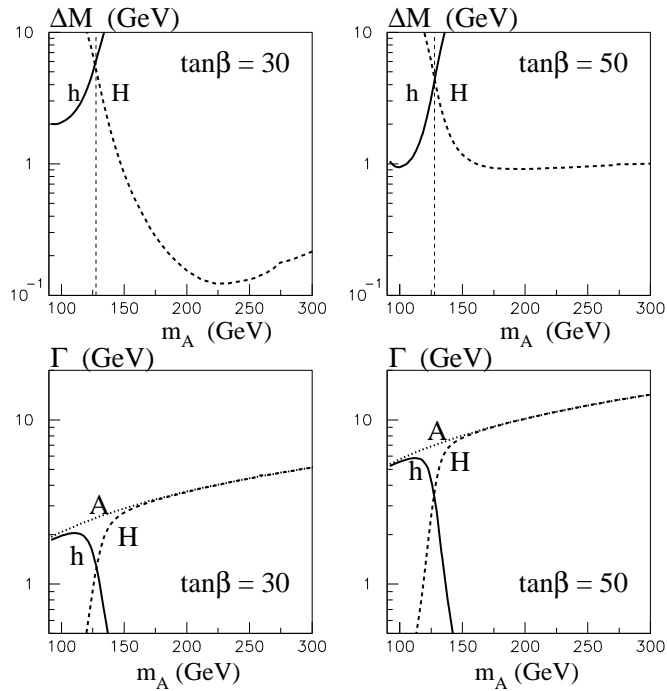}} \caption{The upper plots show the differences $\Delta M = m_A -
m_h$ (continuous curve) and $\Delta M = m_H-m_A$ (dashed curve) as a function of $m_A$ for $\tan\beta = 30$ and 50
respectively.  The lower plots show the total widths of the MSSM neutral high bosons. \label{fig:c}}
\end{center}
\end{figure}
It is especially difficult to disentangle the Higgs bosons in the region around $m_A \sim 130$~GeV, where all
three neutral Higgs states practically overlap~\cite{BDN}. Since the traditional non-diffractive approaches do
not, with the exception of the $\gamma\gamma$ and $\mu\mu$ modes, provide a mass resolution better than
10--20~GeV, all three Higgs bosons will appear as one resonance\footnote{Recall that in this regime the
$\gamma\gamma$ decay mode is hopeless. Moreover, the vector-boson-fusion Higgs production, with the $\bb$ or
$\tau\tau$ decays, does not allow the separation of the $h$ and $H$ peaks due to the poor invariant mass
resolution. As shown in~\cite{BDN}, only the detection of the rare dimuon Higgs decay mode may provide a chance to
separate the Higgs states
in inclusive production. Even this would require that the Higgs mass splitting exceeds at 3--5~GeV.}.

As an example, we give the expected number of CEDP  signal and background events
for ${\cal L} = 30~{\rm fb}^{-1}$ for $m_A = 130$ GeV, using the efficiencies and
angular cuts adopted in Ref.~\cite{DKMOR}.  For $\tan\beta$ = 30(50), the number
of the $A\to \bb$ signal
events, $S_A$ = 0.35(1), while the QCD background is $B = 2.5$ for $\Delta M = 1$ GeV.
For this choice of parameters, $m_h = 122.7(124.4)$ GeV, and the number of the $h\to \bb$
events is $S_h$ = 30(71) with background $B$ = 3.5(3.2).
The corrresponding numbers for the $H$ boson are
$m_H = 134.2(133.5)$ GeV, $S_H =47(124)$ and $B = 2.1$.

An immediate advantage of CEDP, for studying this troublesome region, is that the $A$~contribution is strongly
suppressed, while the $h$ and $H$ states can be well separated ($m_H-m_h\simeq 10$~GeV) given the anticipated
experimental mass resolution of $\Delta M \sim 1$~GeV~\cite{DKMOR}, see Fig.~4.
\begin{figure}[htb]
\begin{center}
\centerline{\epsfxsize=10cm\epsfbox{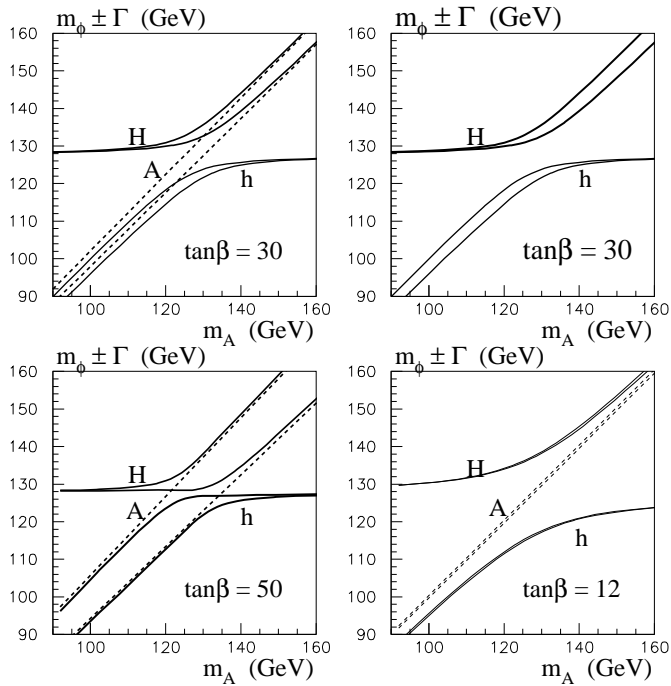}} \caption{The mass bands $m_\phi \pm \Gamma$
for neutral MSSM Higgs bosons as a function of $m_A$, for different values of $\tan\beta$.
A vertical slice through these plots, at a given value of
$m_A$, indicates the position of the MSSM resonance Higgs peaks.  The upper
right hand plot shows that the $h$ and $H$ bosons are clearly identifiable for $\tan\beta=30$, if $A(0^-)$
production is suppressed. The lower plots show how the sensitivity of the widths, to variations of $\tan\beta$,
will change the profile of the peaks. This way of presentation was motivated by Ref.~\cite{BDN}. Note that the
widths of the bosons for $\tan\beta$ less than about 15 become barely visible on this type of plot. \label{fig:d}}
\end{center}
\end{figure}
Note that precision measurements of the two CP-even Higgs states
may allow a direct determination of the
basic MSSM parameter, $\tan\beta$.

Moreover, the forward tagging approach can provide a direct measurement of the width\footnote{Note that, strictly
speaking, when the Higgs width, $\Gamma$, becomes comparable to the mass resolution, $\Delta m$, the zero-width
approximation is not valid any more. Therefore in calculations of the cross sections
the mass resolution and large-width effects should be taken into account.} of the $h$ (for $m_h
\lesim 120$~GeV) and the $H$-boson (for $m_H \gtrsim 130$~GeV). Outside the narrow range $m_A=130\pm 5$~GeV, the
widths of the $h$ and $H$ are quite different (one is always much narrower than the other), see Fig.~3. Recall
that the width measurement can  play an important role in the understanding of Higgs dynamics.  It would be
instructive to observe this phenomenon experimentally.

As discussed in \cite{KKMRCentr}, for $\tan\beta =30$ the central exclusive signal should be still accessible at
the LHC up to an $H$ mass about 250~GeV. For instance, for $m_H=210$~GeV, and LHC luminosity $30~{\rm fb}^{-1}$
($300~{\rm fb}^{-1}$), about 20 (200) $H\to\bb$ events are produced. If the experimental cuts and efficiencies
quoted in {DKMOR} are imposed, then the signal is depleted by about a factor of 6. This leaves 3 (30) observable
events, with background of about 0.1 (1) events.

\section{Calculation of the  $0^\pm$ Higgs cross sections and their uncertainties}

\begin{figure}
\begin{center}
\centerline{\epsfxsize=10cm\epsfbox{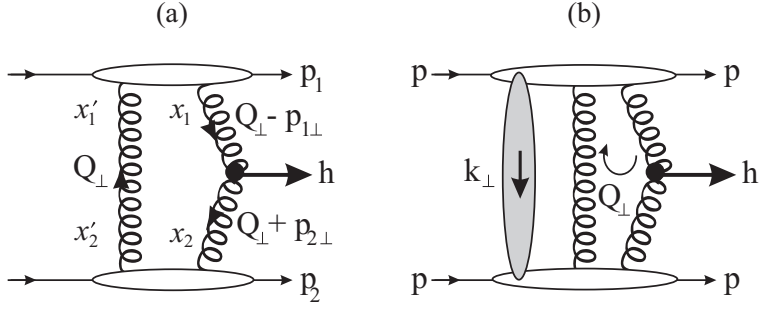}} \caption{(a)~The QCD diagram for double-diffractive exclusive
production of a Higgs boson $h$, $pp\to p + h + p$, where the gluons of the hard subprocess $gg\to h$ are colour
screened by the second $t$-channel gluon.~~(b)~The rescattering or absorptive corrections to $pp\to p + h + p$,
where the shaded region represents the soft $pp$ rescattering corrections, leading to the suppression factor
$S^2$.\label{fig:3}}
\end{center}
\end{figure}

To calculate the cross section for the central exclusive diffractive production of $h(0^\pm)$ Higgs
bosons\footnote{For convenience of presentation in this section we denote the pseudoscalar boson by $h(0^-)$
rather than $A(0^-)$.} we use the formalism of Refs.~\cite{KMR,KMRmm,KMRProsp}.  The amplitudes are described by
the diagram shown in Fig.~\ref{fig:3}(a), where the hard subprocesses $gg\to h(0^\pm)$ are initiated by
gluon--gluon fusion and where the second $t$-channel gluon is needed to screen the colour flow across the rapidity
gap intervals.
Ignoring, for the moment, the screening corrections of Fig.~\ref{fig:3}(b), the Born amplitudes are of the
form~\cite{KKMRCentr}
\be T_h = A\pi^2\int\frac{d^2Q_\perp\ V_h}{Q^2_\perp (\vec Q_\perp - \vec p_{1\perp})^2(\vec Q_\perp + \vec
p_{2\perp})^2}\: f_g(x_1, x_1', Q_1^2, \mu^2; t_1)f_g(x_2,x_2',Q_2^2,\mu^2; t_2), \label{eq:rat3} \ee
where the $gg\to h(0^\pm)$ subprocesses are specified by \cite{KMR,KMRH}
\begin{equation}
A^2 = K\frac{\sqrt{2}G_F}{9\pi^2}\alpha_S^2(m_h^2), \label{eq:Asquared}
\end{equation}\
 with the NLO $K$ factor $K\simeq 1.5$~\cite{MS,KMRProsp}, and by the factors
\be\begin{array}{l l l} V_{h(0^+)} & = & (\vec Q_\perp - \vec p_{1\perp}) \cdot (\vec Q_\perp + \vec p_{2\perp})\nonumber\\
V_{h(0^-)} & = & \left( (\vec Q_\perp - \vec p_{1\perp}) \times (\vec Q_\perp + \vec p_{2\perp})\right)\cdot \vec
n_0\,. \label{eq:rat4}
\end{array} \ee
Here $\vec{n}_0$ is a unit vector in the beam direction.
The $f_g$'s are the skewed unintegrated gluon densities of the proton at the hard scale $\mu$, taken to be
$m_h/2$, with
\be\begin{array}{l l l} Q_1 & = & \min\left\{Q_\perp,|(\vec Q_\perp - \vec p_{1\perp})|\right\},  \nonumber\\
Q_2 & = & \min\left\{Q_\perp,|(\vec Q_\perp + \vec p_{2\perp})|\right\}. \label{eq:rat5}
\end{array} \ee
The longitudinal momentum fractions carried by the gluons satisfy
\begin{equation}
\left(x^\prime\sim\frac{Q_\perp}{\sqrt{s}}\right) \ll \left(x\sim\frac{m_h}{\sqrt{s}}\right) \ll 1
\label{eq:lmf_ineq}
\end{equation}
where, for the LHC, with $\sqrt{s}=14$~TeV, we have $x\sim 0.01$, while $x'\sim 10^{-4}$. Below, we assume
factorization of the unintegrated distributions,
\be f_g(x,x',Q^2,\mu^2;t) = f_g(x,x',Q^2,\mu^2)F_N(t), \label{eq:rat5a} \ee
where we parameterize the form factor of the proton vertex by the form $F_N(t) = \exp(bt)$ with $b=2~\GeV^{-2}$.
In the domain specified by (\ref{eq:lmf_ineq}) the skewed unintegrated densities are given in terms of the
conventional (integrated) densities $g(x,Q_i^2)$. To single log accuracy, we have~\cite{MR01}\footnote
{In the actual computations we use a more precise form as given by eq.(26) of Ref.\cite{MR01}}.
\be f_g(x,x',Q_i^2,\mu^2) = R_g\frac{\partial}{\partial \ln Q_i^2} \left(\sqrt{T(Q_i,\mu)}\: xg(x,Q_i^2)\right),
\label{eq:rat6} \ee
where $T$ is the usual Sudakov form factor which ensures that the gluon remains untouched in the evolution up to
the hard scale $\mu$, so that the rapidity gaps survive. This Sudakov factor $T$ is the result of resumming the
virtual contributions in the DGLAP evolution. It is given by
\begin{equation}
T(Q_\perp,\mu) = \exp\left(-\int_{Q_\perp^2}^{\mu^2}\frac{\alpha_S(k_t^2)}{2\pi}\frac{dk_t^2}{k_t^2}
\int_0^{1-\Delta}\left[zP_{gg}(z) + \sum_q P_{qg}(z)\right]dz\right). \label{eq:T}
\end{equation}
Here we wish to go beyond the collinear approximation and in the $T$ factor to resum, not just the single
collinear logarithms, but the single soft $\log\, 1/(1-z)$ terms as well. Thus we consider the region of
large-angle soft gluon emission, and explicitly calculate the one-loop vertex diagram.  In particular, we do not
approximate $\sin\theta$ by $\theta$.  Then we adjust the upper limit of the $z$ integration in (\ref{eq:T}) to
reproduce the complete one-loop result\footnote {An explicit computation
of the one-loop vertex contribution from the region of gluon
transverse momentum $Q_\perp \ll k_\perp \ll m_h/2$ gives
$(N_c \alpha_S/\pi)[0.212+ {\rm ln}(m_h/2k_\perp)-11/12]dk^2_\perp/k^2_\perp$.
The last number in the square brackets
corresponds to the standard non-logarithmic
contribution related to the $zP_{gg}(z)$ term in (\ref{eq:T}),
while the first number stems from the exact treatment
of large-angle gluon radiation.
This result can be incorporated in (\ref{eq:T})
by simply choosing $\Delta$ as in (\ref{eq:monday1}).}.
We find
\be \Delta = \frac{k_t}{k_t + 0.62m_h}\,. \label{eq:monday1} \ee
The square root in (\ref{eq:rat6}) arises because the bremsstrahlung survival probability $T$ is only relevant to
 hard gluons. $R_g$ is the ratio of the skewed $x'\ll x$ integrated distribution to the conventional diagonal
density $g(x,Q^2)$. For $x\ll 1$ it is completely determined~\cite{SGMR}, with an uncertainty in $R_g$ of the
order of $4x^2$. The apparent infrared divergence of~\eq{rat3} is nullified\footnote{In addition, at LHC energies,
the effective anomalous dimension of the gluon gives an extra suppression of the contribution from the low
$Q_\perp$ domain~\cite{KMRH}.} for $h(0^+)$ production by the Sudakov factors embodied in the gluon densities
$f_g$. However, as discussed in Ref.~\cite{KKMRCentr}, the amplitude for $h(0^-)$ production is much more
sensitive to the infrared contribution. Indeed let us consider the case of small $p_{i\perp}$ of the outgoing
protons. Then, from~\eq{rat4}, we see that $V_{h(0^+)} \sim Q_\perp^2$, whereas $V_{h(0^-)} \sim
p_{1\perp}p_{2\perp}$ (since the linear contribution in $Q_\perp$ vanishes after the angular integration). Thus
the $d^2Q_\perp/Q_\perp^4$ integration for $h(0^+)$ is replaced by $p_{1\perp}p_{2\perp} d^2Q_\perp/Q_\perp^6$ for
$h(0^-)$, and now the Sudakov suppression is not enough to prevent a significant contribution from the
$Q_\perp^2\lesim1~\GeV^2$ domain.

\subsection{Uncertainties in predictions of the cross sections}

To estimate the uncertainty in the predictions for the $h^\pm(0)$ exclusive diffractive cross sections we first
quantify the above uncertainty arising from the infrared region, where the gluon distribution is not well known.
As an example, consider $h(0^-)$ and $h(0^+)$ production at the LHC, for $m_h=120$~GeV and $\mu=m_h/2$, using
different treatments of the infrared region.
We perform calculations using MRST99~\cite{MRST99} and CTEQ6M~\cite{CTEQ} partons respectively with the very low
$Q$ gluon frozen at its value at $Q_{1,2}=1.3$~GeV. Then we integrate down in $Q_\perp$ until $Q_{1,2}$ are close
to $\Lambda_{\rm QCD}$, where the contribution vanishes due to the presence of the $T$-factor. This will slightly
overestimate the cross sections, as for $x\sim0.01$ the gluon density has a positive anomalous
dimension ($xg\sim (Q^2)^\gamma$ with $\gamma > 0$) and decreases with decreasing $Q^2$. A lower
extreme is to remove the contribution below $Q_{1,2}=1.3$~GeV entirely. Even with this extreme choice, the $0^+$
cross section is not changed greatly; it is depleted by about 20\%. On the other hand, as anticipated, for $0^-$
production, the infrared region is much more important and the cut reduces the cross section by a factor of 5.

Another uncertainty is the choice of factorization scale $\mu$. Note that in comparison with previous
calculations~\cite{KMRProsp}, which were done in the limit of proton transverse momenta, $p_{1,2\perp}\ll
Q_\perp$, now we include the explicit $p_\perp$-dependence in the $Q_\perp$-loop integral of (\ref{eq:rat3}). We
resum the `soft' gluon logarithms, $\ln\,1/(1-z)$, in the $T$-factor.
So now the $T$-factor includes both the soft and collinear single logarithms. The only uncertainty is the
non-logarithmic NLO contribution. This may be modelled by changing the factorization scale, $\mu$, which fixes the
maximal $k_t$ of the gluon in the NLO loop correction. As the default we have used $\mu=m_h/2$; that is the
largest $k_t$ allowed in the process with total energy $m_h$. Choosing a lower scale $\mu=m_h/4$ would enlarge the
cross sections by about 30\%. Recall that the general calculation is performed in the collinear approximation.
However, when we take account of large-angle soft gluon emission in the form factor, that is in the calculation of
the virtual-loop correction, as in (\ref{eq:T}) and (\ref{eq:monday1}), we go beyond the collinear approximation.
The corresponding real contribution in (\ref{eq:rat6}) is represented by $R_g$ and the derivative of the $\sqrt T$
factor. Strictly speaking, for the case of large-angle emission there will be some small deviation from this
simplified formula. This gives an extra uncertainty of the order of 10\%.

Next there is some uncertainty in the gluon distribution itself. To evaluate this, we compare predictions obtained
using CTEQ6M~\cite{CTEQ}, MRST99~\cite{MRST99} and MRST02~\cite{MRST02} partons. For $0^+$ production at the LHC,
with $m_h=120$~GeV and $\mu=m_h/2$, we find that the effective gluon--gluon luminosity, before screening, is
\begin{equation}
\left.\frac{d{\cal L}}{dyd\ln m^2}\right|_{y=0} = (2.2,\ 1.7,\ 1.45)\times10^{-2}
\end{equation}
respectively. This spread of values arises because the CTEQ gluon is 7\% higher, and the MRST02 gluon 4\% lower,
than the default MRST99 gluon, in the relevant kinematic region.  The sensitivity to the gluon arises because the
central exclusive diffractive cross section is proportional to the 4th power of the gluon. For $0^-$ production,
the corresponding numbers are $(4.2,\ 2.7,\ 1.7)\times10^{-5}$. Up to now, we have discussed the effective
gluon--gluon luminosity. However, NNLO corrections may occur in the $gg\to h$ fusion vertex. These give an extra
uncertainty of $\pm20\%$. Note that we have already accounted for the NLO corrections for this
vertex~\cite{KMRProsp}.

Furthermore, we need to consider the soft rescattering which leads to a rather small probability, $S^2=0.026$,
that the rapidity gaps survive the soft $pp$ interaction, see Fig.~\ref{fig:3}(b).  From the
analysis~\cite{KMRsoft} of all soft $pp$ data we estimate the accuracy of the prediction for $S^2$ is $\pm50\%$.
One check of the eikonal model calculations of $S^2$ is the estimate of the diffractive dijet production rate
measured by the CDF collaboration \cite{CDFjj} at the Tevatron. The rate, when calculated using factorization and
the diffractive structure functions obtained from HERA data, lies about a factor of 10 above the CDF data.
However, when rescattering corrections are included, and the survival probabilities computed, remarkably good
agreement with the CDF measurements is obtained \cite{KKMR}.

Combining together all these sources of error we find that the prediction for the $0^+$ cross section is uncertain
to a factor of almost 2.5, that is up to almost 2.5, and down almost to $1/2.5$, times the default
value.\footnote{For example, we predict the cross section for the exclusive diffractive production of a Standard
Model Higgs at the LHC, with $m_h = 120$~GeV, to be 2.2~fb with an uncertainty given by the range 0.9--5.5~fb.}
On the other hand, $0^-$ production is uncertain by this factor just from the first (infrared) source of error,
with the remaining errors contributing almost another factor of 2.5.

\section{Simplified formulae for the CEDP signal and background}

Here we give simple approximate formulae for the cross sections for the central exclusive diffractive production
of $h(0^\pm)$ bosons, $pp \to p + h + p$, and for the QCD $\bb$ background as a function of mass.  As was shown in
Ref.~\cite{KMRProsp}, the cross section may be written as the product of the effective gluon--gluon luminosity
${\cal L}$ and the cross section for the hard subprocess $\hat\sigma$
\be \sigma_{\rm CEDP} = {\cal L} \hat\sigma. \label{eq:6a}\ee
At the LHC energy, the luminosity, integrated over rapidity,
may be approximated, to an accuracy of better than 10\% in the interval
$25<m<250$~GeV, by
\be \frac{d{\cal L}}{d\ln m^2} = \int\frac{d{\cal L}}{dy d\ln m^2}\,dy = \left\{
\begin{array}{c l}
{\displaystyle \frac{9.1\times10^5 \, \langle S^2\rangle_{0^+}}{(16 + m)^{3.3}}} & {\rm for}\ h(0^+)\ {\rm production}\\
\,\\ {\displaystyle \frac{1.94 \times 10^3\, \langle S^2\rangle_{0^-}}{(0.5 + m)^{3.45}}} & {\rm for}\ h(0^-)\
{\rm production}
\end{array}
\right. \label{eq:integlumin} \ee
where the mass $m$ is in GeV.  The factors $\langle S^2\rangle$, which are the probability that the rapidity gaps
survive the soft rescattering, are 0.026 and 0.087 for $h(0^+)$ and $h(0^-)$ respectively.  The $0^-$ amplitude
contains the kinematic factor $\vec p_{1\perp}\times \vec p_{2\perp}$, which implies that in impact parameter
space the Fourier transform contains the factor  $\vec b_{1\perp}\times \vec b_{2\perp}$~\cite{KKMRCentr}. Thus
the $0^-$ amplitude tends to populate larger impact parameters, where the suppression caused by soft rescattering
is less effective, than does the $0^+$ amplitude.  The $gg\to h$ subprocesses cross section is, in the zero width
approximation,
\be \hat \sigma_h = \frac{2\pi^2 \Gamma(h\to gg)}{m_h^3}\ \delta \left( 1-\frac{m^2}{m_h^2}\right)\ \sim\
\alpha_S^2(m_h^2)G_F, \ee
where $G_F$ is the Fermi constant.

The $\bb$ background processes were discussed in detail in Ref.~\cite{DKMOR} for a Standard Model $h\to\bb$
signal, with $m_h = 120$~GeV.  There were four appreciable sources of $\bb$ background, each with a
background-to-signal ratio in the range $B/S = 0.06$--0.08, after the appropriate cuts.  In brief, these give
contributions to the background subprocess cross sections of the following approximate forms

\begin{itemize}
\item[(i)] $gg^{PP}\to gg$\quad ($\bb$ mimicked by $gg$): \quad $\hat\sigma\sim \alpha_S^2/m^2$,
\item[(ii)] $gg^{PP}\to \bb$\quad ($J_z=2$ admixture): \quad $\hat\sigma\sim \alpha_s^2/m^2$,
\item[(iii)] $gg^{PP}\to\bb$\quad ($m_b\neq 0$, $J_z=0$ contribution): \quad $\hat\sigma\sim \alpha_S^2m_b^2/m^4$,
\item[(iv)] NLO $gg^{PP}\to \bb g$ contribution: \quad $\hat\sigma\sim \alpha_S^3/m^2$,
\end{itemize}

\noindent where the renormalisation scale is specified by $\alpha_S(m^2/4)$. The $PP$ superscript is to denote
that each incoming gluon to the hard subprocess belongs to a colour-singlet $t$-channel state, see
Fig.~\ref{fig:3}. Note that all background contributions go like $\hat \sigma \sim 1/m^2$, except for (iii). In
addition, for fixed missing mass resolution $\Delta m$, the background contains a factor $\Delta m/m$. Thus for
large $m>100$~GeV the background cross section behaves roughly as
\be \sigma_B = {\cal L} \hat \sigma_B \sim \Delta m/m^6, \ee
while the Higgs signal behaves as
\be \sigma_h = {\cal L} \hat\sigma_h \sim 1/m^3. \ee
To obtain a more precise behaviour one can use (\ref{eq:integlumin}).

In Ref.~\cite{DKMOR} we found a signal-to-background ratio $S/B \sim 3$ for a $m_h = 120$~GeV Standard Model Higgs
boson.  The ratio may, however, be appreciably enhanced for SUSY Higgs with large $m$, and/or large $\tan\beta$.
It is interesting to note that, in the region where the background is larger than the signal ($B>S$), the
statistical significance of the signal $S/\sqrt{S+B}\sim m^{-3}/\sqrt{\Delta m/m^6}$ is practically independent of
mass, assuming that the experimental resolution $\Delta m$ is mass independent.  Moreover, for a low mass, where
$\bb$ pairs in the $J_z=0$ state, (iii), give the dominant contribution to the background, the statistical
significance increases with mass, as here $B\sim \Delta m/m^8$.

Finally, it is important to note that there are still various open questions
which must be addressed by experimentalists in order to establish
the feasibility of CEDP for Higgs searches at the LHC.
One of the most challenging issues concerns triggering.
Assuming that the trigger is based on the central detector
only, the main task is to reduce the rate of triggering jets
in order not to reach saturation of the trigger.
We present below some numbers which may be useful
when considering this issue.

Based on the results of Ref.~\cite{KMRProsp},
we see that the cross section for the  CEDP of
dijets with $E_T > 30$ GeV is 200 pb.
If we assume  that the proton tagging efficiency is 0.6, then
this yields 0.12 events/sec for a luminosity of
$10^{33}~{\rm cm}^{-2}{\rm sec}^{-1}$.
At the same time, the inclusive cross section
for production of a pair of jets, each with $E_T > 30$ GeV  is about
50 $\mu$b.
This leads to 50,000 events/sec, which greatly exceeds
the trigger saturation limit.   For CMS, for example,
this latter limit is 100 events/sec~\cite{EVS}.
Therefore it appears necessary to devise additional topological
requirements in the level-1 trigger for the exclusive signal,
such as making use of the presence of the rapidity gaps.

 \section{Identifying the pseudoscalar boson, $A$}

If the exclusive diffractive cross sections for scalar and pseudoscalar Higgs production were comparable, it would
be possible to separate them readily by the missing mass scan, and by the study of the azimuthal correlations
between the outgoing proton momenta. However, we have seen that the cross section for pseudoscalar Higgs exclusive
production is, unfortunately, strongly suppressed by the $P$-even selection rule, which holds for CEDP at LO. For
values of $\tan\beta \sim 10$--15, the separation between the $H/h$ and $A$ bosons is much larger than their
widths, see Fig.~4(d).  Hence it might be just possible to observe the pseudoscalar in CEDP.  For example, for
$\tan\beta = 15$ and $m_A=95$~GeV, the mass separation, 3.6~GeV, between $h$ and $A$ is about 8 times
larger than the width $\Gamma_h$.  The cross section
\be {\rm Br}(A\to\bb)\cdot \sigma_A \simeq 0.15~{\rm fb}, \ee
when allowing for the large uncertainties in $\sigma_A$, could be just sufficient to bring the process to the edge
of observability.

On the other hand, one may consider the semi-exclusive process
\be pp\to p + j\phi j + p,  \label{eq:sep} \ee
where the Higgs bosons, $\phi=H,A$, are accompanied by two gluon jets with transverse momenta $p_\perp$ satisfying
$m_\phi \gg p_\perp \gg 1$~\GeV.  The emission of soft gluons, with energies (in the $\phi$ rest frame) $E\ll
m_\phi/2$, does not affect the $P$-even selection rule.  Thus in this region the scalar $H$ boson will be
dominantly produced in process (\ref{eq:sep}).

On the other hand, in the domain
\be x_i \equiv 2E_i/m_\phi \sim 1 \ee
we lose the selection rule and have comparable $H$ and $A$ production. Moreover the azimuthal angular
distribution\footnote{The possibility to discriminate the CP-parity of different Higgs states by measuring the
azimuthal correlations of the two tagged gluon jets in inclusive $\phi gg$ production was mentioned in
Ref.~\cite{DKOSZ}. However, the proof of the feasibility of such an approach in non-diffractive $\phi gg$
production requires further study of the possible dilution of the effect caused by parton showers (see
also~\cite{KO}). An exclusive environment of the jets may be needed.} for the gluon jets is different for the $H$
and $A$ cases. In particular, for gluons with $x_1 = x_2 = 1$ accompanying $A$ production, the azimuthal
distribution is practically flat (at LO), whereas for $H$ production it has the form $3+2\cos(2\phi)$. This
observation may be used to distinguish between $H$ and $A$ production, and to study better the properties of the
$A$ boson.

Of course, we have to pay a price, of about a factor $1/20$, to produce two extra jets with $p_\perp$ in the range
$m_\phi\gg p_\perp \gg 1$~GeV, in a rapidity interval with $\Delta \ln x_i\sim1$. However, this is less than the
suppression of $A$ production by the $P$-even selection rule. For $\tan\beta\geq 30$ and $m_A<200$~GeV the cross
section still exceeds 1~fb.

To gain more insight, it is informative to
return to the original central exclusive diffractive process,
which we sketch in Fig.~\ref{fig:3}(a).
In the equivalent gluon approximation, the polarization vectors $\vec\epsilon{ _i}$
 of the gluons, carrying momentum fractions $x_i$, are aligned along
 the corresponding gluon transverse momenta $(Q-p_1)_\perp$
 and $(Q+p_2)_\perp$. For very forward protons
 with $p_{i,\perp }\ll Q_\perp $, this leads to a correlation
 between the polarization vectors, namely $\vec\epsilon {_1}$
is parallel to $\vec\epsilon {_2}$.
 On the other hand, the $\phi $ production vertices behave as
 $V(0^+)\sim \vec\epsilon {_1} \cdot \vec\epsilon {_2}$   and
$V(0^-)\sim \vec\epsilon {_1} \times \vec\epsilon {_2}$.
  Therefore the exclusive diffractive production of the pseudoscalar $A$ boson is
 strongly suppressed\footnote{However, when the polarisation vectors
 $\vec \epsilon{_i}$ are parallel
to $\vec Q_\perp$, it follows, after integration over the azimuthal angle of
 $\vec Q_\perp$, that the LO QCD production of $\bb$ pair
 is also suppressed by a factor $(m_b/E_T)^2$; which results in a very low
 QCD $b\bar{b}$ background to the CEDP.}.

  To change the $t$-channel gluon
 polarization we need particle emission. The best possibility
would be to emit either a
 $\eta'$ or a $\eta_c ~0^-$ meson. Unfortunately the corresponding
 cross sections are too low.
 We lose more than a factor of $10^{-4}$, which is a stronger
suppression than that coming from the $P$-even, $J_z = 0$ selection rule.
For this reason we considered the possibility of emitting a
   (colour singlet) pair of gluons, that is process (\ref{eq:sep}).
  For 'soft' $(x_i\ll 1)$ emission, the dominant contribution comes
 from the term in the triple-gluon vertex which contains
 $g^{\nu\nu'}$,
 and so does not change the gluon polarization vector. Here the indices
$\nu$ and $\nu'$ correspond to the $t$-channel
 gluon polarization before and after the emission of the $s$-channel gluon.
However for
 large $x_i > 1$ the major contribution comes from the term with
 $g^{\nu\mu}$, where $\mu $ is the $s$-channel gluon polarization index. In
 this case after the emission, the new $t$-channel polarization
 vector is directed along the new transverse momentum $Q'_\perp \simeq
 -p_\perp$. Thus for $x_i \sim 1$ the cross sections for $H$ and $A$
 production are comparable, but now the LO QCD
 background is no longer suppressed.

Maybe the best chance to identify the $A(0^-)$ boson is to observe the double-diffractive inclusive process
\be pp  \to    X + \phi + Y,         \label{eq:A1} \ee
where both protons are destroyed. In terms of $t$-channel gluon
 polarization, this is equivalent to the limit $x_i\gg 1$ in the previous
 reaction (\ref{eq:sep}). Process (\ref{eq:A1}) has the advantage of a much
larger cross section\footnote{Note that instead of the factors $1/b$, which came from
the $t_i = -p^2_{i\perp}$
integrals limited by the proton form factors, as in (\ref{eq:rat5a})
(see also eq.(8) in Ref.\cite{KMRProsp}), here we have
logarithmic $dt_i/t_i$ integrals over the intervals
$Q^2_\perp\ll |t_i|\ll m^2_\phi$. Moreover, the Sudakov supression
(which arises from the requirement that no extra
relatively 'hard' gluons are emitted with transverse momenta $k_\perp$ in
the interval
$|\vec{Q}_\perp \pm \vec{p}_{i\perp}|\ll k_\perp \ll m_\phi/2$)
is much weaker than for the exclusive process due to the
larger values of $p_{i\perp}$, see \cite{KMRProsp,KMR}. Therefore the effective
$gg^{PP}$ luminosity for double-diffractive inclusive production is
much larger than in the CEDP case.}. However,
just as for the above
semi-exclusive  process, we do not have the $J_z = 0$ selection rule to
suppress the $b\bar{b}$  background, nor do we have the possibility of
the good missing mass  resolution\footnote{Also, in this
case, pile-up can cause problems.}.  On the other hand, the $gg^{PP}$
luminosity  is more than order of magnitude larger than for the pure
exclusive case. For example, for double-diffractive inclusive
 production, with the rapidity gaps $\Delta \eta >3$, the luminosity is
20 times larger than that for the exclusive diffractive production of a Higgs boson with
mass $m_H=120$ GeV (see for details \cite{KMRProsp}).
So, even for the $\tau\tau$ decay mode we expect a
cross section, $\sigma_{\rm incl}$, of about 20~fb for $A$
production in the MSSM with $\tan\beta=30$ and $m_A=120$~GeV.
 This looks promising, provided that the probability for gluon
misidentification as a $\tau$ is less than 1/150, which appears
to be feasible\cite{CKN,DZEP}.

Despite the QCD background, the observation of the $A$ boson also
appears to be feasible for the $b\bar{b}$ decay
mode, since then $\sigma \sim 200$ fb which gives a
signal-to-background ratio of $S/B=1/24$.
We have evaluated the QCD $b\bar{b}$ background using
the LO formula for the
 $gg^{PP} \to b\bar{b}$ cross section (without any $K$ factor)
 with the polar angle cut $60^\circ<\theta<120^\circ$
as was done in \cite{DKMOR}, and
assumed a mass resolution of $\Delta M=10$ GeV.
The efficiency of tagging the $b$ jets was taken to be about 0.6 \cite{DKMOR}.
 This results in a statistical significance\footnote{If we include a $K$ factor,
$K\sim 1.5$, for the background then the statistical significance is
about $7\sigma$.}
of the signal for the $A$ boson of $S/\sqrt{S+B} \sim 8.5\sigma$, for an
integrated luminosity of 30 fb$^{-1}$.
Moreover one can measure the azimuthal angle $\varphi$ between the
transverse energy $\vec{E}_T$ flows in the forward and backward
regions, that is between the total transverse momenta of the
$X$ and $Y$ systems. The angular dependence is driven by the momenta
transferred across the rapidity gaps, and has the form
\bea &\cos^2 \varphi &\quad {\rm for}\ H(0^+) \nonumber \\
{\rm and} & \sin^2 \varphi &\quad {\rm for}\ A(0^-), \nonumber \eea
whereas for the QCD $\bb$ background the $\varphi$ dependence is flat
(if we neglect the quark mass ($m_b \ll M_{\bb}/2$) and
average over the $b$-jet momentum direction; see  Eq.(15) of
\cite{KMORluminosity} for $Q=0$).
Thus we may suppress the $H(0^+)$ signal by selecting events with $\varphi$ close to
$\pi/2$. We emphasize that here we deal with rather large $E_T$ jets,
$E_T\sim 5-20$ GeV, and so the angular dependences
are largely insensitive to the soft rescatterings.
Furthermore, instead of measuring the $\vec{p}_T$ of the jets (as
in Ref.\cite{DKOSZ}), for process (\ref{eq:A1}) it is enough to measure the total
transverse energy $\vec{E}_T$ in the forward (and backward) regions  of
proton dissociation.

Recall that the effective $gg^{PP}$ luminosity for the inclusive
process (\ref{eq:A1}) was evaluated using the leading order formula.
 Thus we cannot exclude rather large NLO corrections,
particularly for the non-forward BFKL amplitude. On the other hand,
here, we do not enter the infrared domain. Moreover, with
"skewed" kinematics, the NLO BFKL corrections are expected to be
much smaller than in forward case. So we may expect that the uncertainty of
the prediction is about factor of 3 - 4 (or even better).

\section{Conclusion}

The Central Exclusive Diffractive Processes promise a rich physics menu for studying the mechanism of electroweak
symmetry breaking, and the detailed properties of the Higgs sector.  We have seen, within MSSM, that the expected
CEDP cross sections are large enough, especially for large $\tan\beta$, both in the intense coupling and the
decoupling regimes.  Thus CEDP offers a way to cover those regions of MSSM parameter space which may be hard to
access with the conventional (semi) inclusive approaches.  This considerably extends the physics potential of the
LHC and may provide studies which are complementary, both to the traditional non-diffractive approaches at the
LHC, and to the physics program of a future Linear $e^+e^-$ collider. To some extent this approach gives
information which otherwise may have to await a Linear $e^+ e^-$ collider.

\section*{Acknowledgements}

We thank Alfred Bartl, Albert De Roeck, Abdelhak Djouadi, Howie Haber, Leif Lonnblad,
Walter Majerotto, Risto Orava, Andrei Shuvaev, Michael Spira,
and especially Sasha Nikitenko and Georg Weiglein, for useful discussions.  ADM thanks the
University of Canterbury (New Zealand) for an Erskine Fellowship and the Leverhulme Trust for an Emeritus
Fellowship. This work was supported by the UK Particle Physics and Astronomy Research Council, by grants INTAS
00-00366, RFBR 01-02-17383 and 01-02-17095, and by the Federal Program of the Russian Ministry of Industry,
Science and Technology 40.052.1.1.1112 and SS-1124.2003.2.

\end{document}